\documentclass[sigconf, nonacm, natbib=false]{acmart}

\makeatletter
\def\@ACM@checkaffil{
    \if@ACM@instpresent\else
    \ClassWarningNoLine{\@classname}{No institution present for an affiliation}%
    \fi
    \if@ACM@citypresent\else
    \ClassWarningNoLine{\@classname}{No city present for an affiliation}%
    \fi
    \if@ACM@countrypresent\else
        \ClassWarningNoLine{\@classname}{No country present for an affiliation}%
    \fi
}
\makeatother

\usepackage{subcaption}
\usepackage{booktabs}
\usepackage{multirow}
\usepackage{algorithm}
\usepackage{enumitem}
\setlist[itemize]{leftmargin=*} 
\usepackage{algpseudocode}
\usepackage{amsmath}
\usepackage{graphicx}
\usepackage{fontawesome5}
\RequirePackage[datamodel=acmdatamodel, style=acmnumeric,]{biblatex}
\usepackage[bottom]{footmisc}

\usepackage[capitalize]{cleveref}
\crefname{section}{Sec.}{Secs.}
\Crefname{section}{Section}{Sections}
\Crefname{table}{Table}{Tables}
\crefname{table}{Tab.}{Tabs.}
\newcommand{\tool}{\textit{PyTEI}}

\begin{document}
\title{Evaluating and Enhancing Robustness of Deep Recommendation Systems Against Hardware Errors}

\author{Dongning Ma}
\email{dma2@villanova.edu}
\affiliation{%
  \institution{Villanova University}
}

\author{Xun Jiao}
\authornote{Correspondence: Xun Jiao <xjiao@villanova.edu>}
\email{xjiao@villanova.edu}
\affiliation{%
  \institution{Villanova University}
  \institution{Meta Platforms, Inc}
}

\author{Fred Lin}
\email{fred@meta.com}
\affiliation{%
  \institution{Meta Platforms, Inc}
}

\author{Mengshi Zhang}
\email{mengshi@meta.com}
\affiliation{%
  \institution{Meta Platforms, Inc}
}

\author{Alban Desmaison}
\email{alban@meta.com}
\affiliation{%
  \institution{Meta Platforms, Inc}
}

\author{Thomas Sellinger}
\email{tselli@meta.com}
\affiliation{%
  \institution{Meta Platforms, Inc}
}
\author{Daniel Moore}
\email{daniel@meta.com}
\affiliation{%
  \institution{Meta Platforms, Inc}
}

\author{Sriram Sankar}
\email{sriram@meta.com}
\affiliation{%
  \institution{Meta Platforms, Inc}
}

\begin{abstract}
Deep recommendation systems (DRS) are designed to offer customized content with multi-modal contexts such as user profile, interaction history, and item information. 
To handle sophisticated inputs, DRS usually incorporate heterogeneous architectures that may include multi-layer perceptrons (MLPs), embedding tables, and attention mechanisms. 
Large scale DRS heavily depend on specialized high performance computing (HPC) hardware and accelerators to optimize energy, cost efficiency, and recommendation quality. 
However, with the increasing workload of large-scale AI jobs like DRS, modern data center fleets have become more heterogeneous and scaled with a growing number of computing accelerators. This significantly increases the risk of hardware failures which can lead to wrong results and degraded service. Therefore, it is a critical need to study and enhance the robustness of widely deployed AI models with error injection campaign across software and hardware stacks. This paper presents the first systematic study of DRS robustness against hardware errors. We develop \tool, a user-friendly, efficient and flexible error injection framework on top of the widely-used PyTorch framework. \tool ~enables extensive error injection campaign on various DRS models and datasets. We first use dummy models to identify potential factors affecting DRS robustness, and then evaluate 5 realistic models on 3 benchmark datasets. We find that the DRS robustness against hardware errors is influenced by various factors from model parameters to input characteristics. The MLP components inside DRS is particularly vulnerable to hardware errors, and other factors such as the sparse and dense feature ratio and input sparsity also account for the drastic differences of robustness among DRS models and datasets. Additionally, we explore enhancing DRS robustness with 3 error mitigation methods including algorithm based fault tolerance (ABFT), activation clipping and selective bit protection (SBP). Particularly, applying activation clipping can recover up to 30\% of the degraded AUC-ROC score, making it a promising method for enhancing DRS robustness. 

\par \faGithub ~ \url{https://github.com/VU-DETAIL/PyTEI}. 
\end{abstract}

\maketitle
\pagestyle{plain}
\section{Introduction}
\label{sec:intro}
The advances of diverse modern applications including e-commerce, social media, advertising, streaming service and search engine heavily rely on the development of recommendation systems. The objective of a recommendation system is to efficiently provide accurate and high quality personalized recommendations based on the user and item context~\cite{he2017neural, yi2018factorized, zhao2019recommending, ying2018graph}. In response to the increasing scale of data and the tightening quality of service (QoS) requirements, various deep recommendation systems (DRS) emerge, leveraging the outstanding capability of deep learning algorithms~\cite{zhang2019deep, da2020recommendation, naumov2019deep}. Technology companies such as Meta and Amazon use DRS in several ways including news feed, product recommendation, advertisement and marketplace. For example, DRS can leverage a combination of deep learning models to predict what advertisements are most likely to be relevant and engaging to a particular user to improve the quality of advertising.

It is reported that the scale of DRS in production has grown more than an order of magnitude since 2017, urging the use of large-scale high performance computing (HPC) systems to ensure QoS such as tail latency and response time~\cite{lui2021understanding, acun2021understanding, zhao2019aibox, zhao2020distributed}. Recent research has revealed that due to the architectural differences (which is further discussed in detail at \cref{sec:rs_overview}), DRS exhibits drastically different workload characteristics compared with other machine learning systems (e.g. convolutional neural networks (CNN)~\cite{gupta2020architectural, hsia2020cross}), which has thus motivated even the design of specific accelerators (e.g., ASIC) beyond just GPUs~\cite{adnan2022heterogeneous, gupta2021recpipe, guo2021hyperrec,cho2020mcdram}. 

With the increasing workload of large-scale AI jobs like DRS, modern high-performance systems in data centers have become more heterogeneous and scaled with an increasing number of computing accelerators such as GPUs. On the other hand, hardware systems continue to advance in the deep nano-meter regime. Both factors have significantly increased the risk of single hardware failure or data corruption, which can lead to AI job failures. These can take many forms, including radiation-induced soft errors~\cite{hazucha2003neutron}, variation-induced timing errors~\cite{ernst2004razor}, and faults caused by intentional design compromises such as approximate computing and voltage scaling~\cite{han2013approximate}. These faults can result in bit flips during computation, leading to system corruptions, silent data corruption (SDC) or even permanent faults. 

For example, SDC is a type of error that often goes undetected by the system's error detection mechanisms, leading to inaccurate or incorrect results that can have significant consequences for HPC systems and application services. In cloud services, SDC can result in data loss or corruption, causing service disruptions and loss of revenue. Technology companies such as Google and Meta, have reported instances of SDC in their large-scale infrastructure fleet~\cite{bacon2022detection, dixit2022detecting}. Google's study of their production systems found that SDC occurred more frequently than previously believed, with an average rate of a few per several thousand machines~\cite{bacon2022detection}. The existence of SDC in HPC systems with millions of machines can significantly impact application quality, and cause errors or incorrect results toward irremediable loss. More importantly, the traditional data corruption usually leads to noticeable system crashes or data loss, however, SDC can occur without visible symptoms or warnings, making it challenging to detect and diagnose.

In this paper, we propose to systematically evaluate the impact of hardware errors during DRS model inference. We first delve into the architecture of a DRS and perform an analytical study on a benchmark model to evaluate the impact of hardware errors on the output quality. We then sweep across models with different (hyper-)parameters and inputs workload characteristics to investigate the model robustness. To expand the analytical study, we further inject hardware errors into 5 widely acknowledged DRS models with 3 benchmark datasets. We also discuss and examine the effectiveness of 3 error detection methods including algorithm based fault tolerance (ABFT), activation clipping and selective bit protection (SBP). The main contributions of the paper are as follows:
\begin{itemize}
    \item This paper presents the first study of evaluating DRS robustness against hardware errors. Specifically, we systematically explore multiple model hyper-parameters and input characteristics using a dummy model to identify potential factors that can impact DRS model robustness under hardware error. The use of dummy models and synthetic data drastically facilitates the design space exploration process while still providing insightful results to guide further experiments.
    \item We develop a user-friendly, efficient and flexible hardware error injection framework, called \tool, on top of the widely-used PyTorch framework. Using \tool, we conduct an extensive error injection campaign on various DRS models and application datasets. We then inject hardware errors into 5 realistic DRS models on 3 benchmark datasets to evaluate their corresponding robustness. We show that the robustness of DRS models can be affected by factors from both architecture and input features, including \textbf{DRS-specific features} such as dense/sparse feature ratio and input sparsity of data.
    \item We explore enhancing DRS robustness with three error mitigation methods including ABFT, activation clipping and SBP. Particularly, activation clipping can recover up to 30\% of the degraded AUC-ROC score, which paves promising way toward robustness enhancement for DRS.
\end{itemize}

\section{Background}
\label{sec:rs_overview}
Deep learning algorithms are introduced into recommendation systems with the objective of providing high quality and accurate personalized contents to users~\cite{zhang2019deep}. A typical task is the click-through-rate (CTR) prediction, where a DRS is developed to estimate the probability of a user's interaction (click, bookmark, purchase, etc.) with a specific item in a given context~\cite{richardson2007predicting}. Diverse range of DRS have been proposed and implemented. For example, Wide \& Deep (WD) from Google considers the joint advantage of neural networks and factorization models (FM) for enhanced performance~\cite{cheng2016wide}. Further, DeepFM (DFM) enables the learning of both high- and low-order feature interactions as well as a sharing strategy of feature embedding which shows increased performance and efficiency~\cite{guo2017deepfm}. Other experimental and/or commercial architectures such as attentional FM (AFM)~\cite{xiao2017attentional}, deep cross networks (DCN)~\cite{wang2017deep} and deep learning recommendation model (DLRM) from Meta~\cite{naumov2019deep} have also been introduced for higher performance and efficiency. 

\begin{figure}[htbp]
    \centering
    \includegraphics[width = 1\columnwidth]{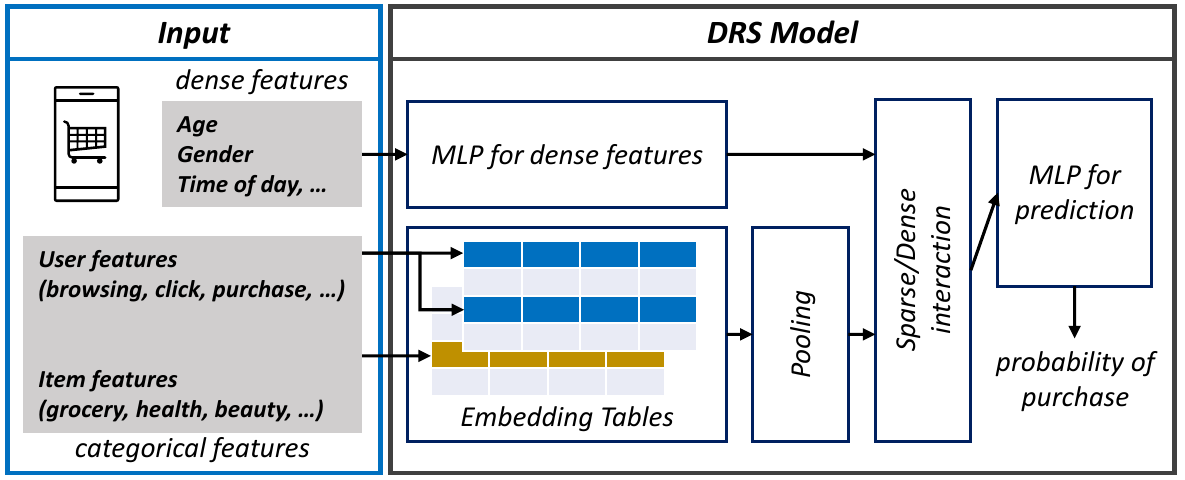}
    \caption{An example DRS model for purchase prediction.}
    \label{fig:drs_overview}
\end{figure}

Without loss of generality, in \cref{fig:drs_overview} we present a typical model of DRS that is trying to predict the purchase probability of an item during a user visit. The input to the DRS model consists of multi-modal features related to the user and the item, where the features can be mainly considered into two groups: the dense features such as the age and gender of the user and the time of day of this visit, and sparse features such as user and item information. Typically, the dense features are passed to MLPs for feature extraction. The sparse categorical features on the other hand, are handled by embedding tables and converted into latent embeddings. After the sparse/dense interaction, the features are then used to predict the probability of user purchase on the item via another MLP. Such probability can be used on a set of items to select relevant candidates to be recommended to the user.

Processing the dense features using neural networks and processing the sparse features with embedding tables have exhibited drastically different workload characteristics during run-time: the neural networks inside an DRS (rendered by MLPs or attentional layers with matrix operations) are usually compute-intensive, while the embedding processes (rendered by indexing embedding tables) are usually memory-intensive~\cite{gupta2020architectural, hsia2020cross}.

\begin{figure*}[htbp]
    \centering
    \includegraphics[width = 1\textwidth]{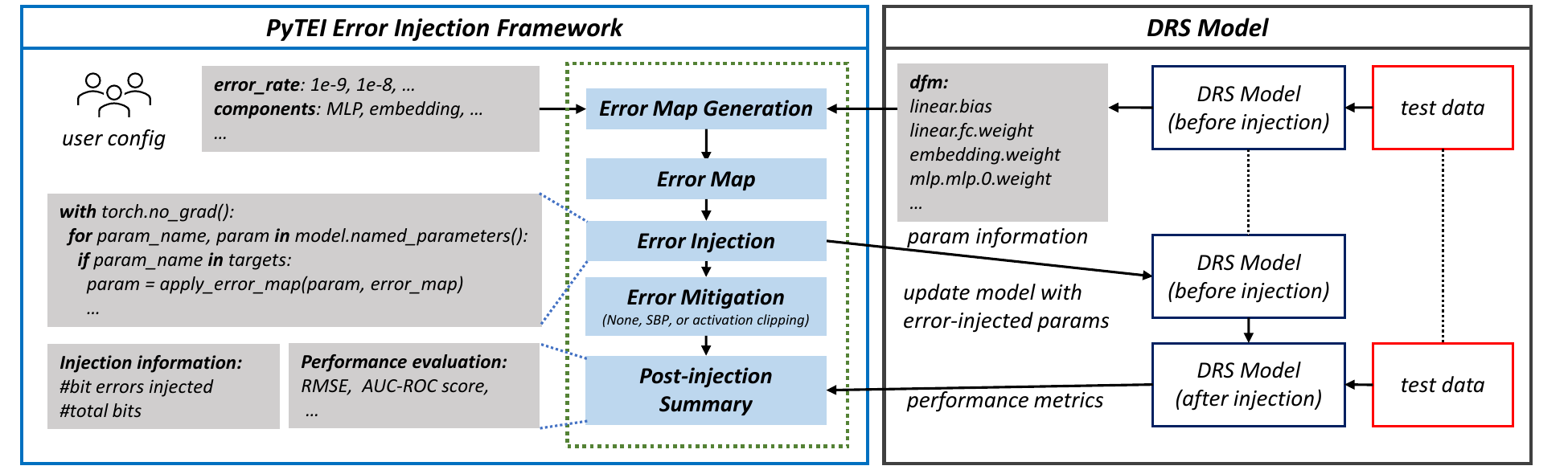}
    \caption{Overview of injecting bit errors in DRS models in PyTorch using \tool.}
    \label{fig:injection_overview}
\end{figure*}

\section{Methodology}
\label{sec:method}
\subsection{Error Model}
In this paper, we focus on the random bit flip error model for its wide acknowledgement to describe hardware errors (e.g., SDCs) in HPC systems~\cite{jiao2020levax, tiwari2015understanding, sullivan2021characterizing, sangchoolie2017one, li2022efficient}. In this model, each bit position can flip based on a probability referred to as bit error rate (BER). The bit flip errors are also statistically independent from each other. Although there are other error models such as permanent errors where the erroneous positions are stuck at a fault value~\cite{sridharan2012study}, we note that they are not the focus of this work since they are less difficult to mitigate compared with random bit flip errors, as permanent errors can be detected and/or mitigated via error correction and detection codes~\cite{sridharan2013feng}. 

\subsection{Error Injection Framework: \tool}
To enable systematic analysis on DRS robustness against SDC, it requires efficient error injection and mitigation so that designs inside the vast design space can be emulated and iterated with reasonable amount of time. We have searched through the available open-sourced error injection frameworks including PyTorchFI~\cite{mahmoud2020pytorchfi}, Ares~\cite{reagen2018ares}, BinFI~\cite{chen2019binfi}, GoldenEye~\cite{mahmoud2022goldeneye} and FIdelity~\cite{he2020fidelity}. However, those frameworks have some common handicaps from being used in this work. The most challenging obstacle is the efficiency. As floating point tensors in PyTorch do not naturally support bit-level operations (such as bit-wise XOR), many frameworks leverage additional packages such as \textit{struct} or \textit{bitstring} for error injection. These packages require frequent conversions between torch tensors and other types of data containers with back-and-forth movements between devices. This accounts for most of the time spent on error injection based on our evaluation, particularly under larger model and higher BERs. Another issue is the tedious setup for environment to use those frameworks. Most of the available frameworks rely on specific environments and/or dockers with numerous additional packages. Some even require users to compile or build CUDA codes by themselves. Moreover, many of them do not provide error mitigation methods or lack of flexible options for hooking in customized error mitigation. As a summary, we need a framework that has the following critical requirements specifically:
\begin{itemize}
    \item The framework should be able to perform bit-level flips on parameters of PyTorch models to inject emulated hardware errors based on a pre-defined BER. The framework should also provide interface to implement error mitigation methods accordingly.
    \item The framework should be easily generalized for models with different architectures or components, convenient to setup and require minimal dependency beyond PyTorch.
    \item The emulation of error injection and mitigation should be considerably fast so that large numbers of different designs can be evaluated efficiently.
\end{itemize}

To realize such requirements, we develop a framework \tool, which is an user-friendly, efficient and flexible hardware error injection framework using no other dependencies besides PyTorch. It has the following key features and contributions which essentially enable us to evaluate DRS robustness at large scale efficiently:
\begin{itemize}
    \item \textbf{User-friendly.} \tool ~does not strictly require any other dependency other than PyTorch. Note that user may use other packages such as \textit{timeit} and \textit{tqdm} for auxiliary purposes such as tracking the progress and time. Additionally, \tool ~also does not require users to compile, build such as other frameworks, nor even use custom environments.
    \item \textbf{Efficient.} \tool ~uses PyTorch built-in data type conversion (\textit{torch.view()}) in bit flip operations, which significantly (about 100X) accelerates the error injection effort. \tool ~can inject bit flip errors at an considerably high BER ($10^{-3}$) to a model with about 19M parameters within a few seconds using even just a commodity CPU. This also enables \tool ~on any PyTorch model as long as they are implemented as \textit{torch.nn.Module} with named parameters.
    \item \textbf{Flexible.} \tool ~provides flexible interfaces for customization on both error model and mitigation. \tool ~allows user to define their customized error model such as random value error beyond the provided bit flip error. \tool ~also implements two mitigation methods for emulation: activation clipping and SBP (introduced in \cref{sec:method_em}) and allows user to customize the mitigation methods.
\end{itemize}

An overview of our approach to inject bit errors into DRS models using \tool ~is present in \cref{fig:injection_overview}. Before error injection, we use a set of test data to obtain golden outputs (from error-free models) and/or test score (e.g. AUC-ROC). In the meanwhile, the user can provide information such as the BER (e.g., 1e-7), the target (e.g., MLP) and parameters (e.g., weights) for error injection. Based on the BER, injection target and model architecture, we generate an error map to indicate what parameters and which bit positions of the parameter to have error. The model and the error map are then iterated together to update the model parameters. With the model after injection, we use the same test data again to observe outputs which is compared with the golden outputs and/or labels to obtain the output deviation and/or the test score degradation.

\section{Evaluation Approach}
\label{sec:eval}
\subsection{Evaluation of DRS Robustness}
\label{sec:exp_method}

\begin{figure*}[htbp]
    \centering
    \includegraphics[width = 0.88\textwidth]{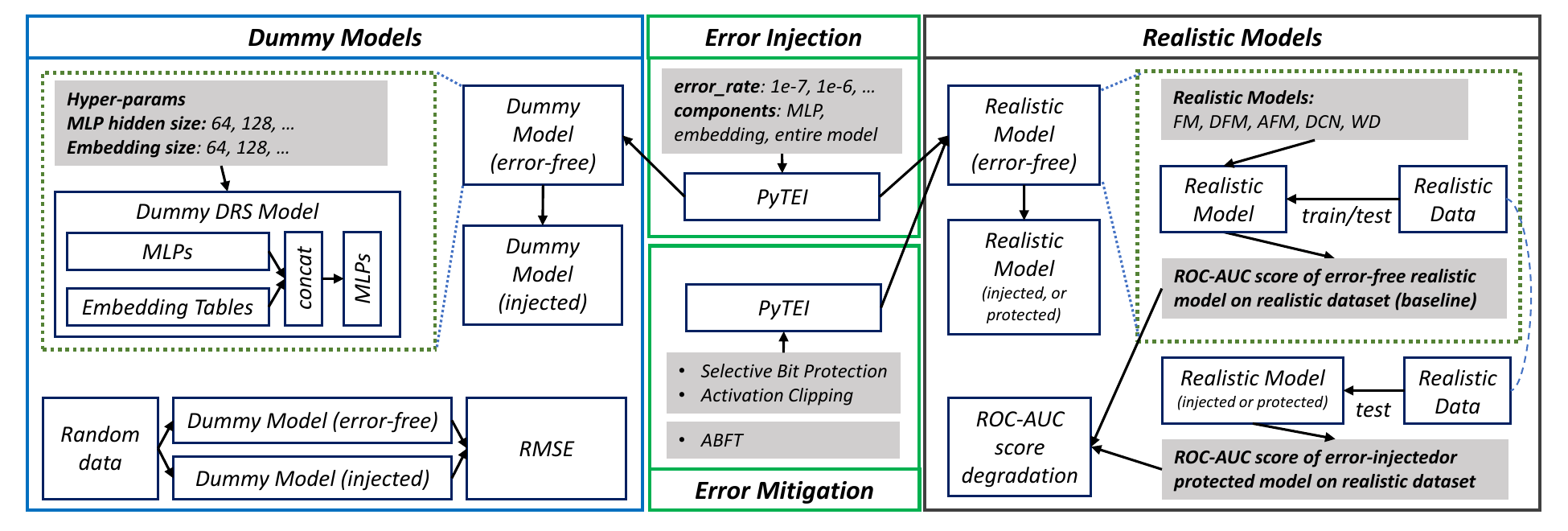}
    \caption{The methodology of experiments: A two-stage robustness evaluation using the dummy model and the realistic models respectively. Each stage features error injection using \tool. Three error mitigation methods are also discussed and/or evaluated in the stage of realistic models.}
    \label{fig:experiment_design}
\end{figure*}

An overview of experiments in this work to evaluate the robustness of DRS against hardware error is shown in \cref{fig:experiment_design}. It features a two-stage evaluation using dummy and realistic models respectively with the error injection framework \tool. The first stage is to inject errors into dummy models, where the objective is to explore a broad design space to provide insights such as identifying the impact of model hyper-parameters and analyzing which of the components inside a DRS are less robust against hardware errors. The second stage is to inject errors into realistic models with the guidance from the observations using the dummy model. The realistic models are trained and tested with realistic datasets as well. Three error mitigation methods are also discussed and/or evaluated for their effectiveness.

For dummy models, we first provide hyper-parameters such as the depth of MLP layers, the size of MLP hidden layer and the size of the latent embedding to configure the model. The model parameters are initialized uniformly~\cite{he2015delving} and \textbf{are not further trained}. The motivation of using dummy models without training is to reduce the effort of model training, which usually takes unrealistic amount of GPU hours (even just by fine-tuning). Additionally, further experiments using realistic models (which are trained) and data align with the observation using the dummy model as discussed in \cref{sec:exp_real}. Therefore we think it is not plausible or necessary to explicitly train each of the model on the entire huge design space. The dummy model is then injected with error to obtain the error-injected model. Random synthetic data are input to both the models and the RMSE score is calculated based on the outputs of the two models to describe the output deviation after error injection which is used to evaluate the hardware error robustness. The detailed results on dummy models are presented in \cref{sec:exp_dummy}.

For realistic models, the architecture is implemented based on corresponding literature. Realistic datasets are used to train and test the model to obtain the baseline scores. After error injection, the same test data are input to the error-injected realistic model to obtain the scores after error injection. The score differences are used to evaluate the DRS robustness. The detailed results on dummy models are presented in \cref{sec:exp_real}. The components inside realistic models can be slightly different from that of dummy models due to architectural differences. For example, the attentional mechanisms inside AFM are also considered as MLP due to their similar implementation using fully connected layers and the major computations inside them are also matrix multiplications.

\subsection{Evaluation of Error Mitigation}
\label{sec:method_em}

In this work, we also present preliminary studies on some error mitigation schemes that are in general implemented for other deep learning models: ABFT~\cite{li2022efficient}, output clipping~\cite{hoang2020ft} and selective bit protection~\cite{chen2018dec}. Analysis and results of evaluation on error mitigation is presented at \cref{sec:exp_em}.

\textbf{ABFT.} ABFT has recently been introduced to DRS for soft error detection~\cite{li2022efficient}. For the general matrix multiply (GEMM) operations like $\mathbf{C} = \mathbf{A} \times \mathbf{B}$, an extra column $\mathbf{S_B}$ is appended to $\mathbf{B}$ of which each element is the corresponding row-sum, i.e., $\mathbf{S_B}[i] = \sum_{j = 0}^{n - 1}\mathbf{B}[i][j]$. Therefore, with the appended column, we can obtain the equality check vector $\mathbf{A}\mathbf{S_B}$ of which each element \textbf{should be} the corresponding row-sum of $\mathbf{C}$. Therefore, if such equality is not hold, an error is detected. The result of this GEMM operation will be discarded and the operation is also repeated. For embedding operations, an additional column $\mathbf{C_T}$ is similarly added alongside embedding table $\mathbf{T}$ of which each element is the corresponding row-sum as well. Assume indices $\mathbf{F}$ are hit during an embedding operation, then the output embedding should be $\mathbf{R} =  \sum_{f \in F} \mathbf{T}[f]$. The same indices are also used to index the elements in the additional column to obtain the equality check value $r = \sum_{f \in F} \mathbf{C_T}[f]$ and the value $r$ \textbf{should be} equal to the sum of $\mathbf{R}$. Similarly, if such equality is not hold, an error is detected.

\textbf{Activation Clipping.} Clipping the network activations has been investigated to improve the fault tolerance~\cite{hoang2020ft}, or enable ultra-low precision designs~\cite{choi2018pact}. For example, the ReLU activation function in the model is revised by \cref{eq:lo_clip}~\cite{hoang2020ft} or \cref{eq:lo_clamp}~\cite{choi2018pact} where activation will be confined within a range. $T$ is the pre-defined threshold for clipping.

\begin{equation}
  f(x) = 
  \begin{cases}
    x & \text{if } 0 \leq x \leq T \\
    0 & \text{otherwise}\\
  \end{cases}
  \label{eq:lo_clip}
\end{equation}

\begin{equation}
  f(x) = 
  \begin{cases}
    0 & \text{if } x < 0\\
    x & \text{if } 0 \leq x \leq T \\
    T & \text{if } x > T \\
  \end{cases}
  \label{eq:lo_clamp}
\end{equation}

\textbf{Selective Bit Protection.} Preliminary studies have illustrated that certain bit positions in specific data types may have dominating effect on the error behavior~\cite{chen2018dec}. For example, in IEEE-754 floating point data format, the value of a 32-bit float number is determined by 3 fields of bits: the sign-bit, the 8-bit exponent filed and the 23-bit mantissa field as shown in \cref{eq:ieee754}. Therefore, the most significant bits in the exponent field are observed to have the most critical contribution of the value and is prioritized to have protection from bit errors~\cite{chen2018dec} when it is impossible to protect all the bit positions due to resource limitations. In \cref{sec:exp_em}, we provide analytical and/or experimental results on those error mitigation schemes and compare their effectiveness and efficiency. 

\begin{equation}
    \text{val} = \text{sign} \times 2^{\text{exponent}} \times \text{mantissa}
  \label{eq:ieee754}
\end{equation}

For a concise summary of our methodology, we aim to explore and answer the following research questions in this work:
\begin{itemize}
    \item \textbf{RQ1}: How will DRS models generally be impacted by hardware errors? What components inside a DRS model are more vulnerable, or vice versa?
    \item \textbf{RQ2}: How will hyper-parameters such as the hidden layer size of MLPs and the embedding dimensions affect the robustness against hardware errors?
    \item \textbf{RQ3}: Will error mitigation methods for other deep learning models such as ABFT, activation clipping and SBP still be effective for DRS models? 
\end{itemize}

\section{Experiments: Dummy Models}
\label{sec:exp_dummy}
\subsection{Dummy Models and Data Generation}
For simplicity, we build a dummy DRS model with minimal configuration. The inputs to the model are straightforwardly synthesized, each of which has two vectors representing the dense and sparse features, respectively. To ensure simplicity, the dense features are floating point numbers generated with standard Gaussian distribution, and the sparse features are binary and in Bernoulli given a probability (sparsity). The models consists of a MLP handling the dense features, an embedding table to handle the sparse features and another MLP as the predictor, as shown in \cref{fig:experiment_design}. The hyper-parameters of the dummy model and the parameters for generating the dataset which establish the design space to explore according to \ref{tab:dummy_dse}. Specifically, we explore the impact of the depth and hidden layer size of the MLP model, the dimension of embedding for models and the input sparsity and sweep the BER from $10^{-9}$ to $10^{-2}$. We show below that such dummy models with synthetic data have already shown remarkable performance in identifying DRS robustness vulnerability.

\begin{table}[htbp]
\small
  \centering
  \caption{Hyper-parameters of the dummy model, the parameters for generating the input and the evaluated BERs.}
    \begin{tabular}{|c|c|c|}
    \hline
    \multirow{3}{*}{model hyper-parameters} & MLP depth & 1, 2 \\
    \cline{2-3}          & MLP hidden layer size & 64, 128, 256, 512 \\
    \cline{2-3}          & embedding dimension & 64, 128, 256, 512 \\
    \hline
    \multirow{3}{*}{input characteristics} & dense dimension & 128 \\
    \cline{2-3}          & sparse dimension & 8192 \\
    \cline{2-3}          & sparsity & 0.001, 0.01 \\
    \hline
    \multicolumn{2}{|c|}{bit error rate} & from 1e-9 to 1e-2 \\
    \hline
    \end{tabular}%
  \label{tab:dummy_dse}%
\end{table}%

\subsection{Robustness Analysis of Dummy Models}
\begin{figure*}[htbp]
    \centering
    \includegraphics[width = .97\textwidth]{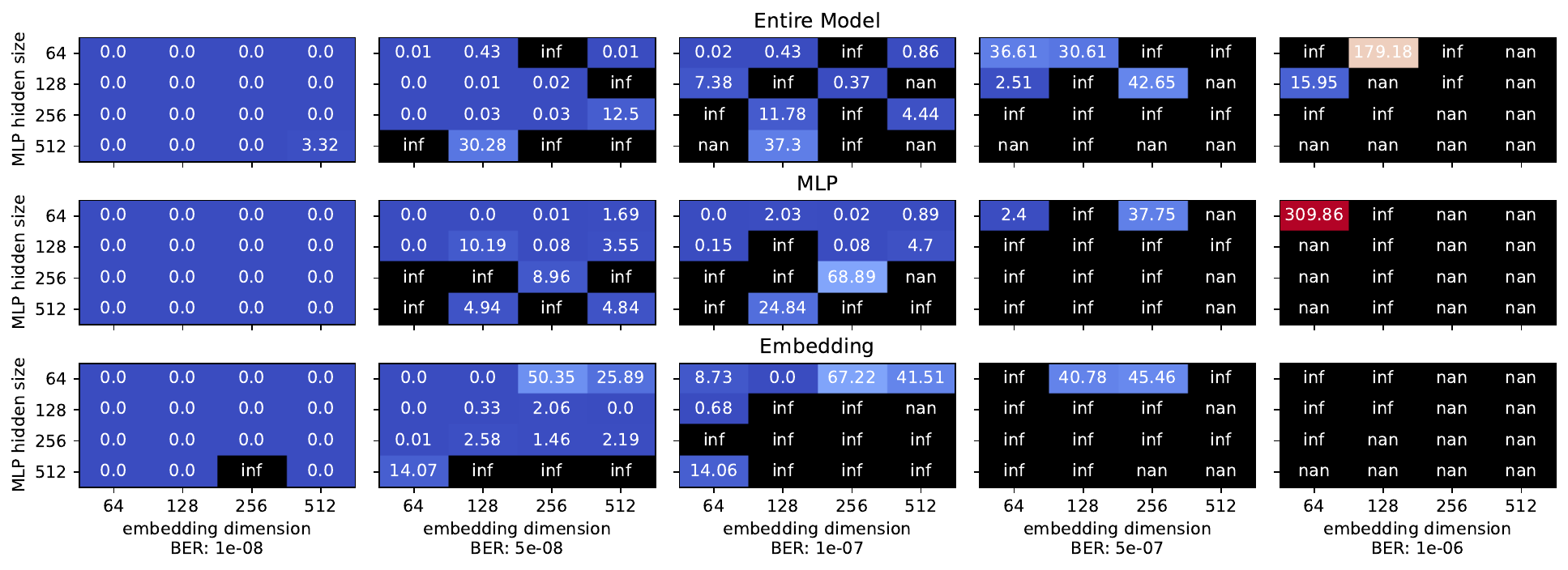}
    \caption{Root mean square error (RMSE) between outputs of error-free models and outputs from error-injected models under different \textbf{BERs}, \textbf{MLP hidden layer size} and \textbf{embedding dimensions}. Errors are injected into different components inside the model as outlined by the title of each row.}
    \label{fig:heatmaps_0}
\end{figure*}

\begin{figure*}[htbp]
    \centering
    \includegraphics[width = .97\textwidth]{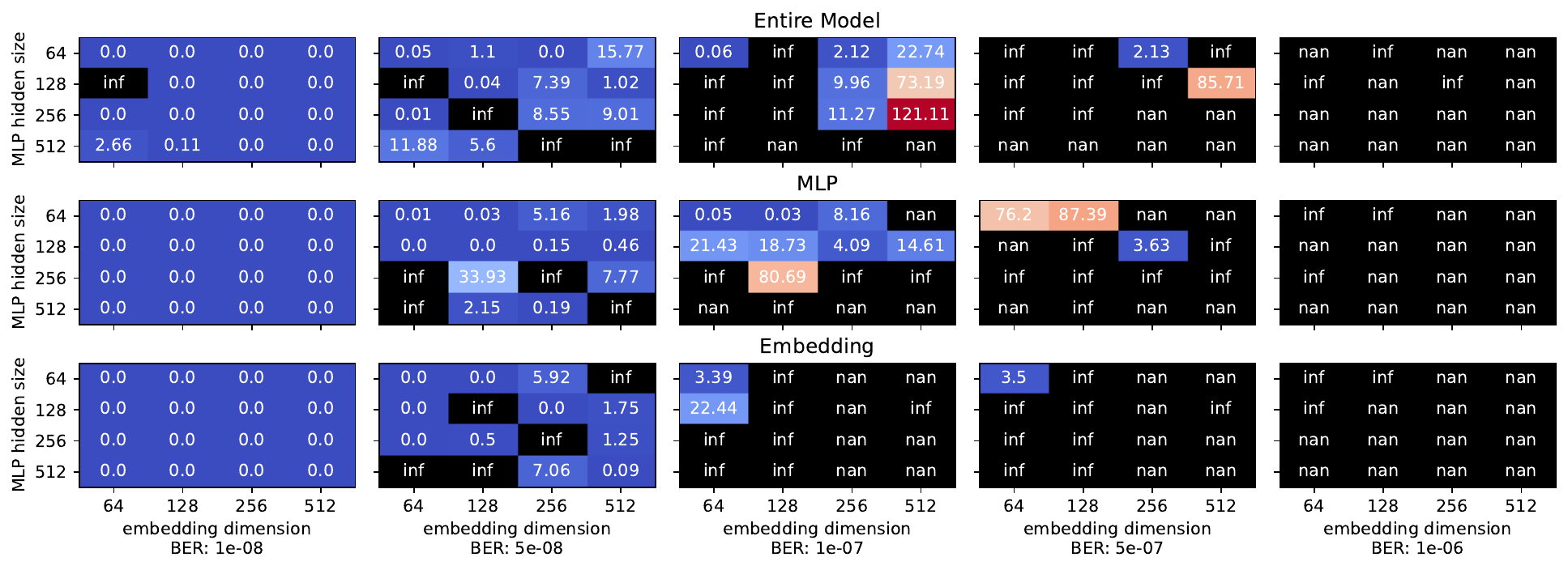}
    \caption{Root mean square error (RMSE) between outputs of error-free models and outputs from error-injected models under different \textbf{BERs}, \textbf{MLP hidden layer size} and \textbf{embedding dimensions} \textit{when the number of hidden layers inside MLP is increased from 1 to 2}. Errors are injected into different components inside the model as outlined by the title of each row.}
    \label{fig:heatmaps_1}
\end{figure*}

\begin{figure*}[htbp]
    \centering
    \includegraphics[width = .97\textwidth]{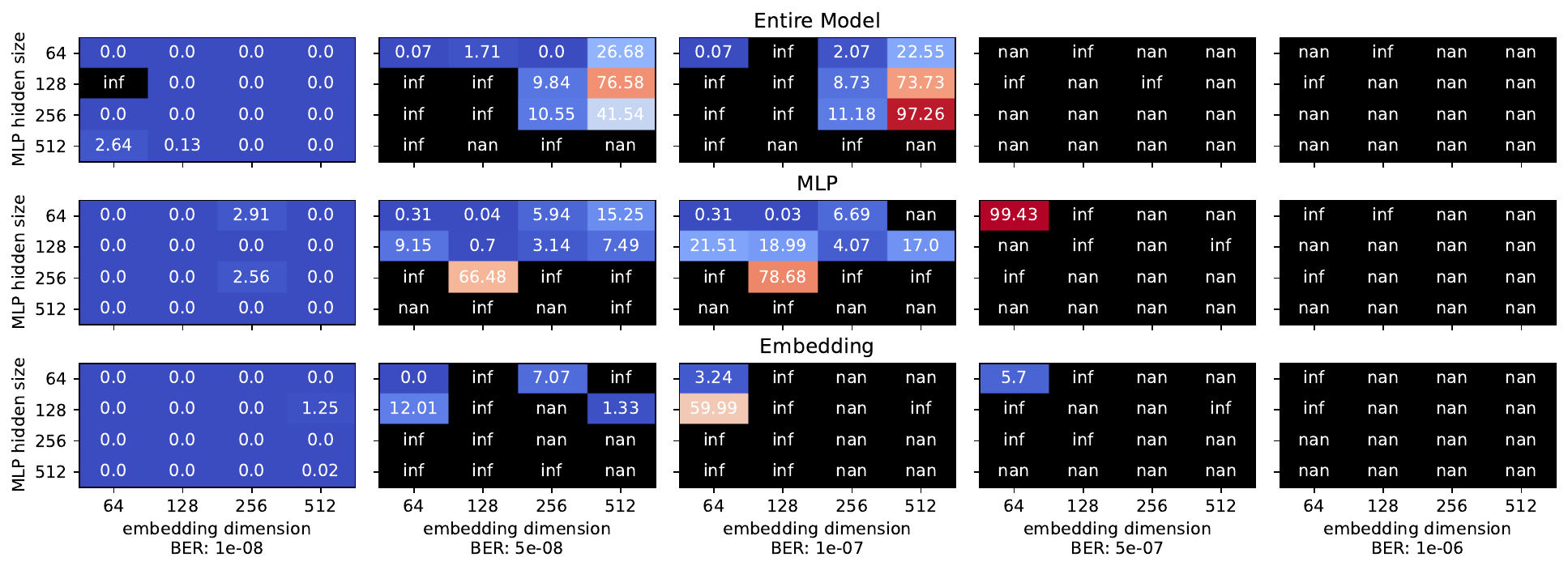}
    \caption{Root mean square error (RMSE) between outputs of error-free models and outputs from error-injected models under different \textbf{BERs}, \textbf{MLP hidden layer size} and \textbf{embedding dimensions} \textit{when the number of hidden layers inside MLP is increased from 1 to 2 and the input sparsity is increased by 10X to 0.01}. Errors are injected into different components inside the model as outlined by the title of each row.}
    \label{fig:heatmaps_1s}
\end{figure*}

As briefly described in \cref{sec:exp_method} and \cref{fig:experiment_design}, for dummy models, we use the RMSE between the outputs from error-free and error-injected models to evaluate the deviation and then characterize the impact of hardware errors. From \cref{fig:heatmaps_0} to \cref{fig:heatmaps_1s} we show the RMSE metrics of the all the design points explored. We omit the RMSEs for BERs below $10^{-8}$ and above $10^{-6}$ because they are either in all zeros or all \textit{inf} or \textit{nan} numbers. According to our observation, there are possibilities that the bit errors happen on critical positions within a number such as the exponent field in the IEEE-754 32-bit floating point number~\cite{kahan1996ieee} and result in extremely large deviations (e.g., flipping the 2nd bit of the number $0.625$ will get $2.13 \times 10^{38}$) which, after network propagation, can exceed the range of floats and become \textit{inf} or \textit{nan}. Therefore, if there are more than 1 such invalid numbers of the output, we mark the RMSE as \textit{inf} or \textit{nan} accordingly in the figure. On the other hand, we mark any RMSE less than 0.005 as 0.0 for presenting in the figure, indicating there is no noticeable errors observed. For example, the 0.0 at the upper left slot of the row ``Entire Model'' indicates that when both the \textbf{MLP hidden size} and the \textbf{embedding dimension} are 64, injecting errors to the weights on the \textbf{entire model} does not induce in any noticeable error on the output. 

We can make several observations from the figure. Overall for the figures from \cref{fig:heatmaps_0} to \cref{fig:heatmaps_1s}, when the BER increases the RMSE metric will also increase accordingly. This is intuitive and aligns with the observations of other machine learning models in general, since more errors injected into the model will likely translate into higher output deviations. However, we do observe that tuning the hyper-parameters of different components inside a DRS can uncover the different characteristics of robustness. For example, although by increasing the MLP hidden size (from top to down in each sub-figure) and the embedding dimension (from left to right in each sub-figure) can both lead to increasing error, MLP hidden size seems to induce more impact by having higher RMSE, or more invalid numbers. 

Based on the results we infer that such difference between MLP and embedding originates from their corresponding architecture. For MLP, even with drop-out schemes, each weight as a parameter will more or less contribute when calculating prediction results with the forward pass. Therefore, when increasing the hidden layer size of MLP, the outputs also become more influenced. However, as to embedding, for each sample only a limited number of entries in the embedding table are indexed due to the sparsity of inputs, therefore it is of a high chance that the indexed entries are not affected by the error. Thus, increasing the embedding dimension does not exacerbate the output error as significant as MLP.

\textbf{Impact of MLP depth.} In \cref{fig:heatmaps_1}, we show the results when increasing the depth of the MLPs in the DRS model by adding another hidden layer. It can be observed that the added hidden layer deteriorates the output quality regardless of whether the error injection is into MLP, embedding or the entire model. According to our analysis, the reason behind of the increasing error is two-fold: first, an additional hidden layer in MLP indicates more parameters. Thus, with the same BER the absolute number of bit errors is also higher. Second, if the bit error in the first hidden layer result in an erroneous value, it can be propagated to the second layer and cause more erroneous values due to the fully connection. 

\textbf{Impact of input sparsity.} In \cref{fig:heatmaps_1s}, we further change the sparsity of the sparse vectors in the input set by $10X$ from 0.001 to 0.01. By comparing the last row ``Embedding'' in \cref{fig:heatmaps_1s} and \cref{fig:heatmaps_1}, we can observe that since more entries are indexed within the embedding table, more errors are incorporated in the latent embeddings. Then the impact from error-injected embedding table will subsequently become more pronounced, which is presented by having higher RMSE and/or more invalid numbers. 

As a summary, from the experimental results on dummy models we can have the following insights: First, similar to other machine learning models, DRS model suffers a similar trend that when the BER increases, the output quality will degrade in general. However, the hyper-parameters of different components inside a DRS can pose different impact on the robustness. This is resulted from the inherent architectural differences of those components. When the BER increases to around $10^{-6}$, nearly all the results witness invalid values in their outputs. Therefore, we can infer that for the experiments on realistic model and data, the AUC-ROC score will experience noticeable degradation for BERs above $10^{-6}$.

\section{Experiments: Realistic Models}
\label{sec:exp_real}

\subsection{Realistic Models and Datasets}
We target at 5 widely acknowledged recommendation systems: Factorization Machine (FM)~\cite{rendle2010factorization}, Deep Factorization Machine (DFM)~\cite{guo2017deepfm}, Attentional Factorization Machine (AFM)~\cite{xiao2017attentional}, Deep Cross Network (DCN)~\cite{wang2017deep}, and Wide and Deep (WD)~\cite{cheng2016wide}. The models are implemented from \textit{torchfm}\footnote[1]{\url{https://github.com/rixwew/pytorch-fm}} and the details about the evaluated models are listed in \cref{tab:models}. As it is very difficult to find pre-trained PyTorch models for those DRS models since different models use different frameworks for implementations and datasets for training, we train all the models afresh using the default parameters in \textit{torchfm}. 

We use AUC-ROC score as our quality metric for consistency on comparison since it is used mostly across the papers of the DRS models and use early stopper that the training terminates when the AUC-ROC score does not improve for 3 consecutive epochs. The baseline AUC-ROCs reported in \cref{tab:models} mostly align with what the original research papers present though we do note that they can still be slightly different. Since the main objective of this paper is not focused on competing the scores but on evaluating the robustness, we recognize such difference acceptable for research in this work.

We use 3 benchmark datasets for CTR prediction: Movielens-1M\footnote[2]{\url{https://grouplens.org/datasets/movielens}}, Movielens-20M and Criteo\footnote[3]{\url{https://ailab.criteo.com/ressources}} DAC dataset. The Movielens-1M dataset contains 1 million ratings from 6,000 users on 4,000 movies and the Movielens-20M dataset contains 20 million ratings and 465,000 tag applications applied to 27,000 movies by 138,000 users. The Criteo dataset contains the click records of 45 million users with 13 continuous and 26 categorical features. The datasets are randomly split by 8:1:1 for training, validation and testing. 
\begin{table}[htbp]
\small
  \centering
  \caption{Models, Datasets and Baseline AUC-ROC Scores}
    \begin{tabular}{cccc}
    \toprule
    \textbf{Model} & \textbf{Dataset} & \textbf{\#Parameters} & \textbf{Baseline AUC-ROC} \\
    \midrule
    \multirow{3}[2]{*}{fm~\cite{rendle2010factorization}} & MovieLens-1M & 171K  & 0.813 \\
          & MovieLens-20M & 4.59M & 0.838 \\
          & Criteo & 18.5M & 0.779 \\
    \midrule
    \multirow{3}[2]{*}{dfm~\cite{guo2017deepfm}} & MovieLens-1M & 170K  & 0.842 \\
          & MovieLens-20M & 4.59M & 0.838 \\
          & Criteo & 18.5M & 0.785 \\
    \midrule
    \multirow{3}[2]{*}{afm~\cite{xiao2017attentional}} & MovieLens-1M & 170K  & 0.862 \\
          & MovieLens-20M & 4.58M & 0.827 \\
          & Criteo & 18.5M & 0.801 \\
    \midrule
    \multirow{3}[2]{*}{dcn~\cite{wang2017deep}} & MovieLens-1M & 161K  & 0.824 \\
          & MovieLens-20M & 4.32M & 0.818 \\
          & Criteo & 17.4M & 0.799 \\
    \midrule
    \multirow{3}[2]{*}{wd~\cite{cheng2016wide}} & MovieLens-1M & 170K  & 0.828 \\
          & MovieLens-20M & 4.59M & 0.815 \\
          & Criteo & 18.5M & 0.791 \\
    \bottomrule
    \end{tabular}%
  \label{tab:models}%
\end{table}%

\subsection{Robustness Analysis of Realistic Models}
\begin{figure*}[htbp]
    \centering
    \includegraphics[width = 0.93\textwidth]{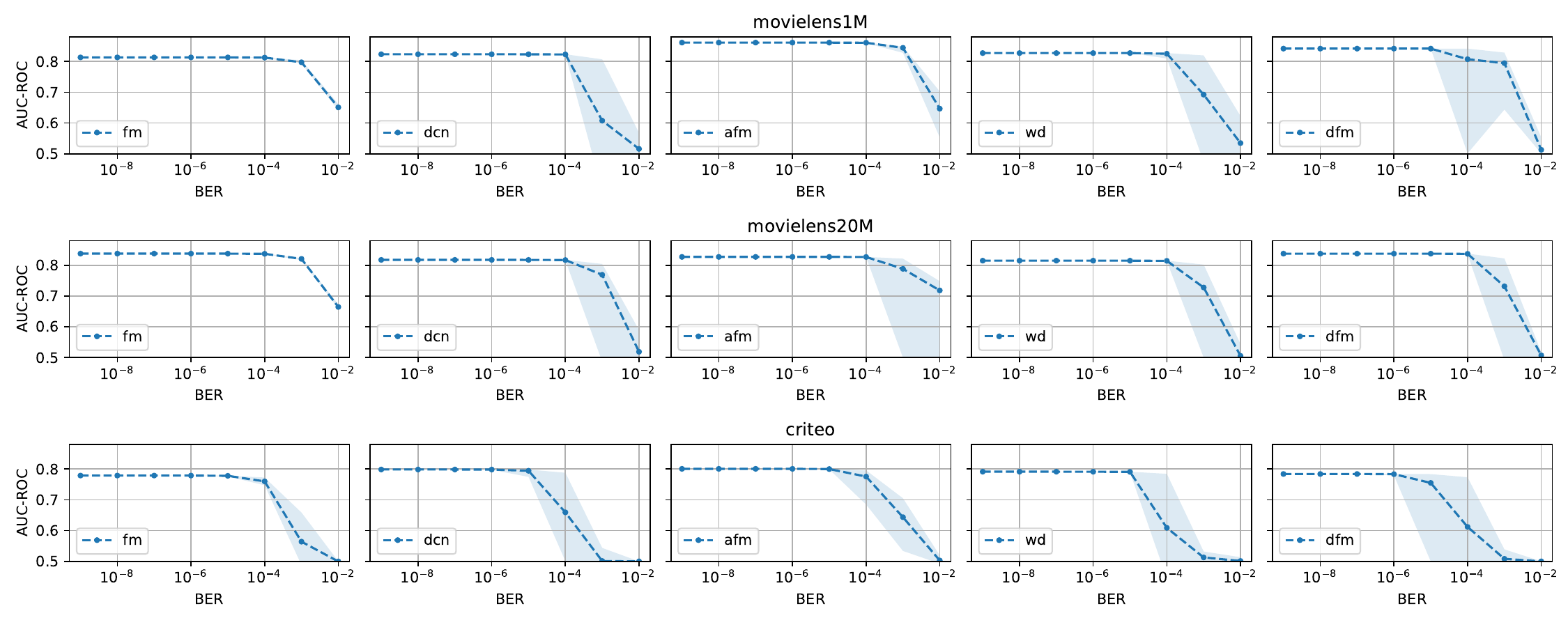}
    \caption{Performance of recommendation systems under different bit error rates. Errors are injected into the weights in \textit{both embedding tables and neural networks} in each model. Results are averaged from 10 different runs of error injection with shaded areas covering the range of maximum and minimum scores observed.}
    \label{fig:performance_ber}
\end{figure*}

\begin{figure*}[htbp]
    \centering
    \includegraphics[width = .93\textwidth]{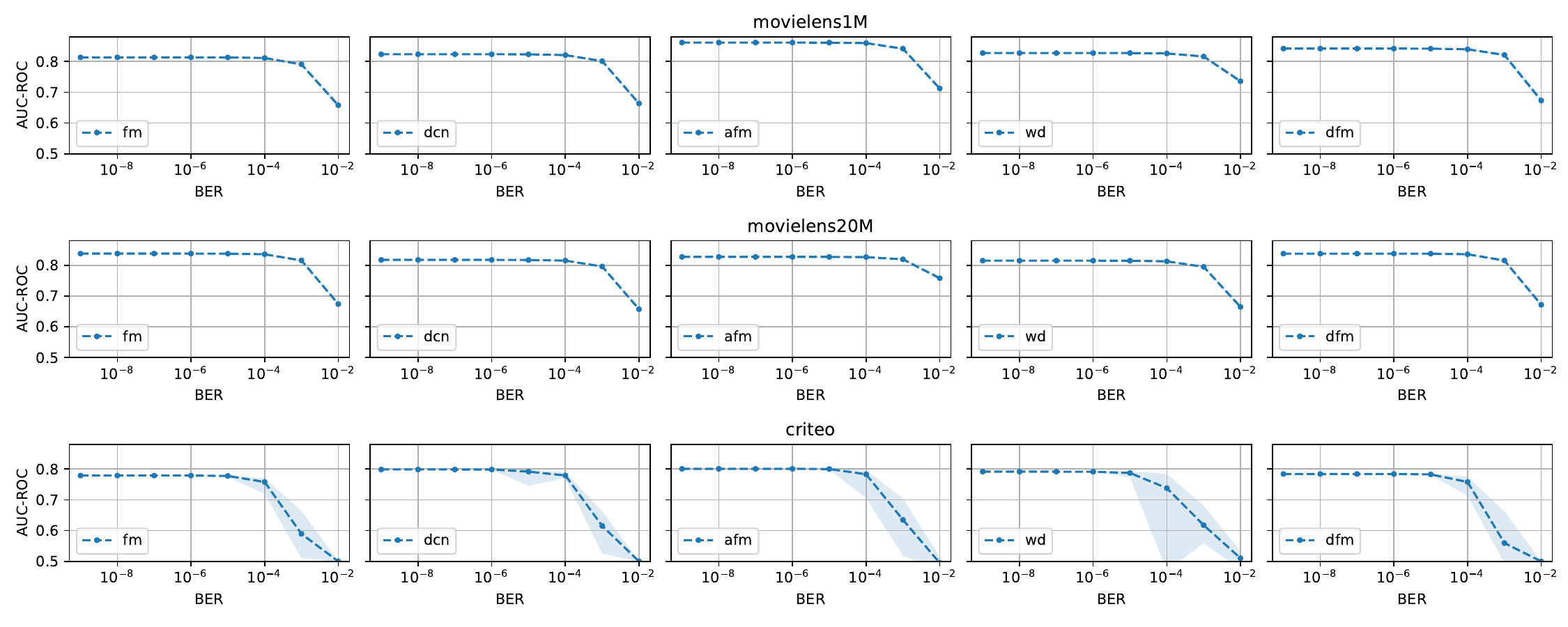}
    \caption{Performance of recommendation systems under different bit error rates. Errors are injected into the weights in the \textit{embedding tables}. Results are averaged from 10 different runs of error injection with shaded areas covering the range of maximum and minimum scores observed.}
    \label{fig:performance_ber_embed}
\end{figure*}

\begin{figure*}[htbp]
    \centering
    \includegraphics[width = .93\textwidth]{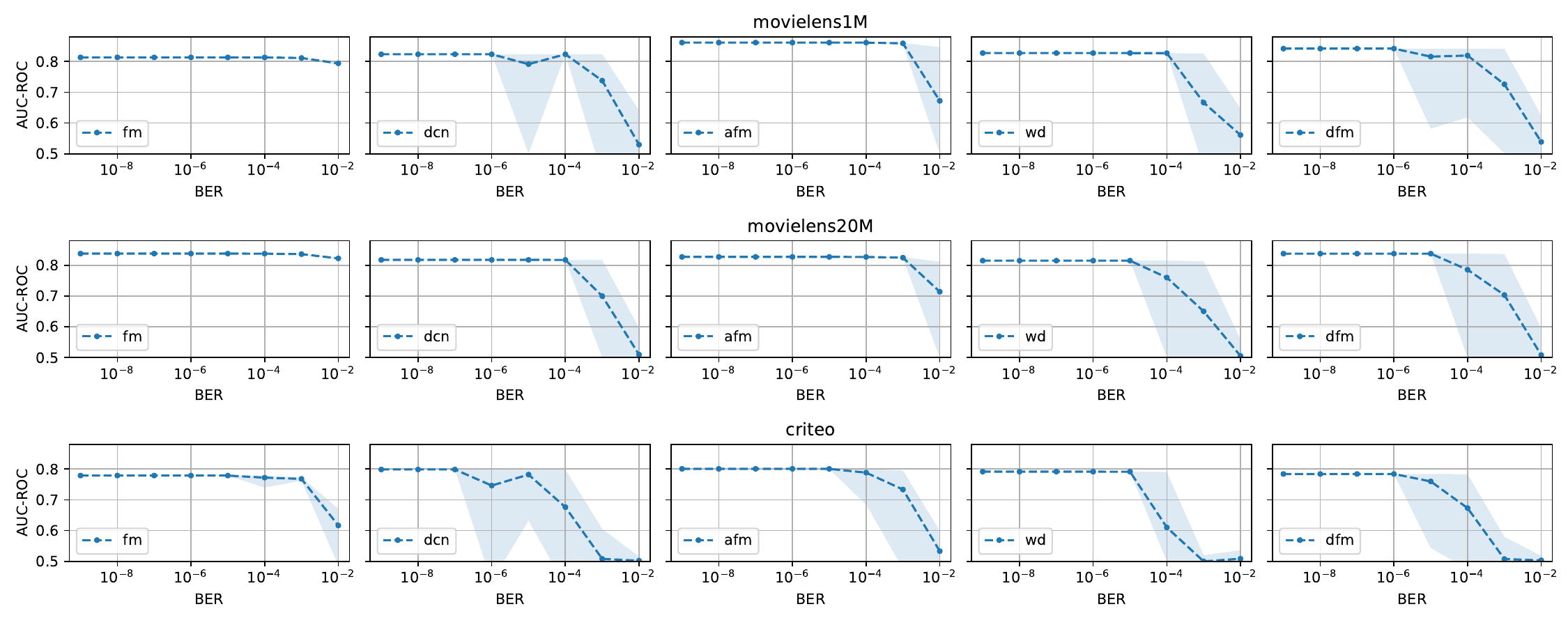}
    \caption{Performance of recommendation systems under different bit error rates. Errors are injected into the weights in the \textit{neural networks} in each model. Results are averaged from 10 different runs of error injection with shaded areas covering the range of maximum and minimum scores observed.}
    \label{fig:performance_ber_nn}
\end{figure*}

In \cref{fig:performance_ber} to \cref{fig:performance_ber_nn} we present the AUC-ROCs score of the evaluated realistic DRS models under different BERs from $10^{-9}$ to $10^{-2}$. The line shows the average of the AUC-ROC scores while the shade represents the range of the scores observed from 10 individual experiments. We can first observe that the general trend is consistent with what we have observed from experiments on dummy models: with higher BER, the model outputs (scores) suffers from higher quality degradation. Moreover, for most of the models, the scores exhibit noticeable quality loss after the BER of $10^{-6}$, which also aligns with our inference using dummy models. However, severe score drops start from $10^{-5}$ or $10^{-4}$ and most of the models becomes completely disrupted into random guess (AUC-ROC score of around 0.5) with BER higher than $10^{-3}$.

By using dummy models we observe that different components inside the DRS model have different robustness characteristics. We also have similar observations from realistic models: the embedding tables are more robust against error as the significant error drops start after BER $10^{-3}$ as shown in \cref{fig:performance_ber_embed}. The range of scores of 10 runs is also more consolidated near the average by having smaller shading areas. Moreover, models with more MLP architectures such as DCN, DFM and WD are less robust against hardware errors compared with embedding-dominant architectures such as FM (which only has MLP for the final predictor), based on \cref{fig:performance_ber_nn}. 

We have also observed that different datasets can also impact the model robustness. For dummy models, we use randomly synthesized data with fixed dimensions of dense and sparse features as well as the sparsity. However, it can be observed that models developed for the Criteo dataset are less robust against errors, as the BER where they exhibit noticeable score degradation is lower than that of the two MovieLens datasets. This can result from dataset-specific attributions, for example the models developed for Criteo have in general higher amount of MLP architectures compared with the rest datasets. The Criteo dataset has less sparsity which also influences the robustness of DRS model as analyzed using dummy models.

As a summary, from the experimental results on realistic models and datasets we can have the following insights: In general, the robustness of DRS against hardware error intuitively follows the trend of other deep learning models, i.e., the score gradually decreases when BER increases. However, the robustness of DRS drastically (up to 2 orders of magnitude) vary between different models and datasets. Such huge variation, based on our analysis and observation, originates from two major aspects: the architectural differences and the characteristics of input data. A DRS model with more MLP architectures, higher amount of dense input features and less sparsity in sparse input features is likely to be more susceptible against hardware errors and therefore exhibits lower robustness and requires more intensive error mitigation.

\section{Experiments: Error Mitigation}
\label{sec:exp_em}

\textbf{ABFT.} ABFT is proved to be effective to detect soft errors in DRS of having 99\% effectiveness and 10\% false positive of bit flip error detection with overhead below 26\% on various DRS architectures~\cite{li2022efficient}. However, most of such error correction methods are capable of single-error correcting and double-error detecting (SEC-DED), and simply re-execute when an error is detected but not correctable. When we attempt to implement ABFT, the experiment hardly finishes and outputs any meaningful results. Particularly when the BER is higher than $10^{-4}$, almost all the matrices for parameters under such higher BER contain multiple errors which always force re-execution of operations under ABFT scheme. Therefore, we conclude that such error-correction-code based methods work for intermittent, low-rate soft and transient errors caused by, e.g., cosmic rays and radiation effects yet are infeasible for scenarios with higher BERs which can come from logic and data-path errors, or voltage and frequency scaling.
\begin{figure*}[htbp]
    \centering
    \includegraphics[width = 0.95\textwidth]{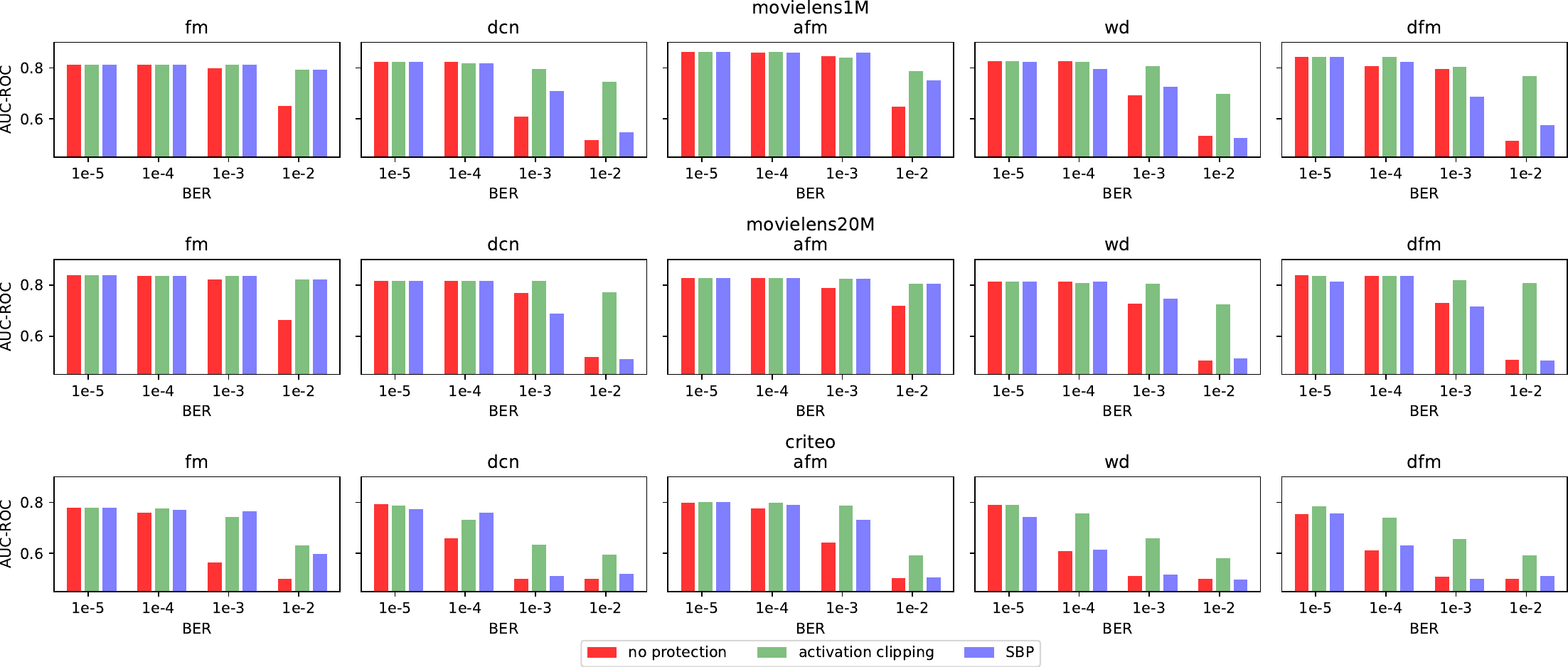}
    \caption{Comparison between no protection, activation clipping, and selective bit protection under different BERs. Results are averaged from 10 individual runs.}
    \label{fig:mitigation}
\end{figure*}

\textbf{Activation Clipping and SBP.} For activation clipping we manually clip the activations inside the range of $[-6, 6]$ (inspired by ReLU6~\cite{howard2017mobilenets}). For SBP, we protect the sign and exponent bits. Although this work, which aims to provide primary evaluation on the error mitigation methods for DRS, does not focus on exploring the optimal parameters for activation clipping or the bit positions to protect for specific DRS models, we provide functionality for users to customize those parameters in \tool. We present the results of error mitigation using activation clipping and SBP in \cref{fig:mitigation}. The BERs start from $10^{-5}$ since any lower BER does not incur significant degradation in score. We can observe that mostly, both of the error mitigation schemes are effective to regain some AUC-ROC score across all the DRS models and all the datasets by 0.05 to 0.3.

However, activation clipping compared with SBP is more effective on regaining the score loss, particularly for models that have deep MLP architectures such as DCN, WD and DFM. The effectiveness of error mitigation also varies across different datasets. Generally, models that show stronger robustness (FM and AFM) as discussed in \cref{sec:exp_real} also experience better recovery. Those observations align with our analysis that the MLP components inside DRS models, when having error, contribute to most of the output quality degradation. Thus, clipping activation from MLPs can enable better error recovery compared with generally protecting selective bit positions of all the parameters. Another advantage of activation clipping is its smaller overhead. Protecting selective bit positions require dedicated mechanisms or hardware. Clipping activation of MLP layers are less resource-limiting since it can be simply implemented by modifying the activation functions.

\section{Discussion}
In this section, we can discuss the research questions as mentioned at the end of \cref{sec:method}:
  
\textbf{RQ1:} How will DRS models generally be impacted by hardware errors? What components inside a DRS model are more vulnerable to hardware errors, or vice versa?

\textbf{A1:} Similar to other machine learning models, DRS models are also negatively impacted by hardware errors. However, different components inside a DRS exhibit different hardware error robustness due to the drastic heterogeneity of architectures incorporated. According to observations from \cref{sec:exp_dummy} and \cref{sec:exp_method}, MLP architectures are usually considered less robust while embedding tables are more robust against hardware errors. Therefore, we recommend future system designers to prioritize MLPs for protection to mitigate the impact of hardware errors.

\textbf{RQ2:} How will hyper-parameters such as the hidden layer size of MLPs and the embedding dimensions affect the robustness against hardware errors?

\textbf{A2:} Tuning the hyper-parameters will influence the hardware error robustness of DRS models. Similarly, different components inside DRS also exhibit different sensitivity on hyper-parameter tuning. In general, by increasing the hidden layer size of MLPs in DRS, the robustness suffers from more severe degradation comparing with increasing the embedding dimension. Additionally, dense/sparse feature ratio and input sparsity are the unique robustness characteristics of DRS compared with other deep learning models, which are originated from the architectural difference between MLP and embedding components, as analyzed in \cref{sec:exp_dummy}. 

\textbf{RQ3:} Will error mitigation methods for other deep learning models such as ABFT, activation clipping and SBP still be effective for DRS models?

\textbf{A3:} According to our analysis and observation, conventional ABFT is implausible to enhance the robustness of DRS under our error scenario due to their theoretical handicap. Activation clipping and SBP are effective for the evaluated DRS models to recover degraded score. However, activation clipping is observed to be more effective than SBP since it specifically protects the MLP components inside the DRS model which are less robust against hardware errors based on our analysis. Activation clipping also requires less overhead compared with SBP.

\section{Related Works}
Hardware errors can originate from various sources. Various beam testings have revealed that cosmic rays and radiation can induce soft transient errors in GPU memories~\cite{tiwari2015understanding, sullivan2021characterizing}. Additionally, manufacturing defects and device aging can induce permanent faults which are particularly common in deep learning accelerators with systolic arrays~\cite{zhang2018analyzing, hanif2020dependable, zhao2022fsa}. In addition to faults, hardware errors can also be intentional, e.g., from the trade-off between accuracy and efficiency which is acknowledged as approximate computing~\cite{mittal2016survey, xu2015approximate}. For example, voltage scaling can induce logic or timing errors in memories~\cite{yuksel2021mors, yang2017sram, yauglikcci2022understanding} or computational units~\cite{tziantzioulis2015b, jiao2020levax}, which can degrade the deep learning model output quality. Approximate or inexact logic in circuit design can also be applied to functional units such as multipliers for accuracy and power trade-offs~\cite{chen2020optimally, mrazek2016design, shafique2016cross}. Due to the diverse sources of error, the corresponding error rate can also drastically vary, which motivates us to evaluate across a broad range of BERs to accurately assess deep learning model robustness in this paper.

There are method to mitigate the impact of hardware errors in deep learning systems to enhance their robustness. For example, limiting the activation has been explored to improve the resilience of hardware faults for DNNs by confining the output of activation functions within a valid range~\cite{hoang2020ft}. For specific data types, critical bits protection is also investigated to mitigate the impact of hardware bit flip errors~\cite{chen2018dec}. Error code correction is also commonly used to identify and mitigate the impact of hardware errors, e.g., single bit and double bit error correction using ECC requires about 12.5\% redundancy for HBM2 memory on compute-class GPUs~\cite{chen2018configurable}. ABFT has been recently evaluated on the MLP and embedding components of a DRS model on soft errors~\cite{li2022efficient}, but it assumes the error correction to be simply re-execution of the operation impacted, which becomes implausible under our error schemes (as detailed in \cref{sec:exp_em}). In addition, there are other frameworks for fault or resilience analysis such as Thales~\cite{tyagi2022thales}, and FIdelity~\cite{he2020fidelity} which consider hardware factors such as flip-flop reuse, but those frameworks still largely focus on single-bit soft and/or transient errors. Such absence motivates us to present results with realistic models and datasets in this paper.

\section{Conclusion}
With burgeoning deployment of DRS on HPC with various architectures from GPUs to ASICs, the associated hardware errors from diverse sources become a more pronounced concern to threat the QoS. In this paper, we systematically analyze the robustness of DRS against SDC from hardware errors and evaluate the factors that significantly contribute to the degraded output quality under hardware error. We develop \tool, a user-friendly, efficient and flexible framework for error injection. We identify that the MLP components inside DRS models are the most susceptible to hardware errors, while other major factors such as the input sparsity account for the drastic robustness difference between DRS models for different datasets. We also provide evaluation on 3 error mitigation methods including ABFT, activation clipping and SBP. Activation clipping, based on our observation, is the most promising error mitigation method which can regain up to 0.3 AUC-ROC score with minimal overhead. This paper provides a systematic effort on evaluating hardware error robustness the emerging DRS models, providing insights to the promising future research directions.

\printbibliography
\end{document}